\documentclass{aa} 

\usepackage{graphicx}
\usepackage{textcomp, gensymb}
\usepackage{booktabs}
\usepackage{tablefootnote}
\usepackage[flushleft]{threeparttable}
\usepackage{xcolor}
\usepackage[varg]{txfonts}
\begin{document} 

   \title{Active galactic nucleus feedback in NGC 3982}

   \author{Prajwel Joseph,\inst{1,2}
          Koshy George,\inst{3}
          K. T. Paul\inst{1}
          }

   \institute{Department of Physics, CHRIST (Deemed to be University), 
              Bangalore 560029, India\\
              \email{prajwel.pj@gmail.com}
              \and
              Indian Institute of Astrophysics, Bangalore 560034, India
             \and
             Faculty of Physics, Ludwig-Maximilians-Universität, Scheinerstr. 1, Munich 81679, Germany
             }

   \date{Received January 15, 2000; accepted March 16, 2000}

 
\abstract
{The energetic feedback from supermassive 
black holes can influence star formation 
at the centres of galaxies. 
Observational evidence for active galactic nucleus (AGN) impact on 
star formation can be searched 
for in galaxies by combining ultraviolet imaging 
and optical integral field unit data. 
The ultraviolet flux directly traces 
recent star formation, and the integral 
field unit data can reveal dust attenuation, 
gas ionisation mechanisms, and gas  
kinematics from the central regions of the 
galaxy disk. 
A pilot study on NGC 3982 shows star formation 
suppression in the central regions of 
the galaxy, likely due to negative AGN feedback, 
and enhanced star formation in the outer regions.
The case of NGC 3982 could be observational 
evidence of AGN feedback operating in a Seyfert 
galaxy.}


   \maketitle
%

\section{Introduction}

Galaxies in the local Universe host supermassive 
black holes (SMBHs) with masses $>$ 10$^6$ M$_\odot$ 
at their centres, which can accrete large amounts
of gas from the immediate vicinity \citep{2013ARA&A..51..511K}. 
The accretion changes the SMBH and the 
host galaxy to an active galactic nucleus 
(AGN) phase with a net effect 
of energy release in the form of radiation, jets,
and outflows. 
The energy output can heat the cold molecular 
gas and increase the turbulence, suppressing 
star formation and regulating the gas accretion 
onto the SMBH. 
This self-regulating process is termed 
AGN feedback. It is believed to be an 
important aspect of galaxy formation and 
evolution, with various pieces of observational 
evidence supporting it as a star formation 
quenching pathway for galaxies
(see \citealt{fabian2012observational} 
and \citealt{morganti2017many} for reviews).
Active galactic nucleus feedback is included in galaxy 
simulations to match observed galaxy properties 
\citep{springel2005black, 
       di2005energy, 
       somerville2008semi, beckmann2017cosmic}. 
While it may be anticipated that this feedback 
results in the suppression of star formation
(negative AGN feedback) due to the copious amount
of energy released by the AGN, 
there is also the inverse case of positive feedback,
where star formation is triggered due to 
 AGN activity \citep{zinn2013active}.
 
Although evidence for 
AGN feedback is available, direct observations 
of its influence on star formation in 
the local Universe remain limited.
One reason for the scarcity 
of direct observations could be that AGN 
luminosities vary considerably during a typical 
star-forming episode, and the effect 
of feedback on star formation may not be readily 
apparent \citep{hickox2014black}. 
Such variability can give rise to 
situations where even if a galaxy's star 
formation were affected by AGN activity, 
the corresponding signatures
of a strong AGN could have become undetectable.
Similarities in star formation efficiencies
between Seyfert and inactive local galaxies have 
been found by \cite{rosario2018llama}; while this
observation may suggest that the AGN 
activity does not affect 
star formation efficiencies in AGN 
host galaxies, it is also possible
that the AGN could have been active in the
past in inactive galaxies.
In studies where the sample of galaxies 
is selected by AGN activity, it may 
be challenging to obtain a correlation between 
the highly variable activity of AGN feedback and star 
formation that takes place on a relatively long 
timescale.

However, there may exist galaxies that 
hosted AGN in the recent past, where we can 
directly probe the evidence of AGN activity 
affecting star formation.
If we can find such galaxies and study 
recent star formation in the central regions 
and its connection to any recent AGN activity, 
it may lead to a better understanding of the 
complex relationship of AGN feedback with 
star formation. 
Such studies might make it possible to find 
relatively smoothly time-variable parameters 
connected to AGN feedback and explore 
their relationship with star formation as a proxy 
to unravel how AGN feedback transforms the galaxy. 


Among the observations of AGN feedback 
affecting star formation in nearby galaxies, 
there are cases of both positive and negative feedback.
The AGN jet-induced positive 
feedback is found in Centaurus A, NGC 1275, 
and Minkowski's Object 
\citep{mould2000jetCenA, 
       ngc1275canning2010star, 
       van1985minkowski}.
Outflow-induced star formation exists
near the nuclear region of NGC 5643
\citep{cresci2015ngc5643}.
In NGC 7252 and the jellyfish galaxy JO201,
AGN activity is proposed to have
suppressed star formation in 
the central region \cite{george2018uvit, george2019gasp}.
Both positive and negative feedback has been
observed in NGC 5728 \citep{shin2019positive}.
It will be interesting to increase the statistics of galaxies with evidence of AGN feedback on star 
formation present in the 
local Universe.

A comprehensive search for observational 
signatures of AGN feedback on star formation 
in nearby galaxies could help unravel the 
complex spatial and temporal relationship of 
AGN interaction with its host galaxy environment.
 Since AGN are known to impart energy
through radiative (sometimes called quasar or wind)
and mechanical modes (also known as radio, kinetic, or jet; \citealt{harrison2017impact}),
they leave behind observational signatures
that could be identified. 
The gas around AGN will be ionised
by the large energy throughput and 
produce excitation lines that can be spatially mapped. 
Similarly, star formation can be 
suppressed or triggered in the galaxy,
which can be observed via 
the absence or presence of emitted ultraviolet 
(UV) flux associated with star formation activity.

Integral field unit (IFU) based spectroscopy of a galaxy disk allows us to spatially resolve the excitation 
mechanisms present in the galaxy;  
this is particularly useful in identifying 
the extent of AGN-excited regions.
The UV imaging data directly probe recent star 
formation (< 200 Myr, \citealt{kennicutt2012star}), 
and any effect of AGN feedback
on the galaxy should be revealed in
UV images as reduced flux due to suppressed star formation.  
Integral field unit observations of nearby galaxies 
from Mapping Nearby Galaxies at APO (MaNGA) 
and UV data from the Galaxy Evolution Explorer (GALEX)
can be used to check 
for possible spatial evidence of AGN feedback 
\citep{martin2005galaxy, bundy2014overview, gunn20062, drory2015manga, smee2013multi}.

Our objective is to study the 
impact of AGN activity on star formation 
using MaNGA and GALEX data. 
We created a sample of galaxies with the
Sloan Digital Sky Survey (SDSS) optical 
spectral classifications\footnote{\url{https://www.sdss.org/dr17/spectro/catalogs/\#Objectinformation}} 
of `AGN' or `BROADLINE'.  
We removed edge-on galaxies from our sample 
following a visual check of SDSS \textit{urz} 
imaging data and selected only those galaxies
with both MaNGA and GALEX data.
There are 86 galaxies in the sample.
NGC 3982 is the nearest face-on galaxy 
in the sample and was selected for our 
pilot study, where we demonstrate the 
feasibility of our project in identifying 
the effect of AGN activity on star formation.
NGC 3982\footnote{\url{http://simbad.u-strasbg.fr/simbad/sim-id?Ident=NGC\%203982}} 
(UGC 6918) is a late-type galaxy at z = 0.00371\footnote{\url{http://skyserver.sdss.org/dr17/VisualTools/explore/summary?sid=1146264831163131904}}.
NGC 3982 is classified as a Seyfert 1.9 type 
galaxy based on optical spectra \citep{veron2010catalogue}.
Very-long-baseline interferometry (VLBI) 
observations of NGC 3982 in 1.7 and 5 GHz 
reveal that there could be jet or outflow structures 
on milliarcsecond scales \citep{bontempi2012physical}.
\cite{gonzalez1993star} noted the presence of 
circum-nuclear star formation in the galaxy.
We note that 1 arcsecond in the sky 
corresponds to 0.076 kpc 
at a distance of 15.6 Mpc \citep{wright2006cosmology}.

The paper is organised as follows.
Section 2 describes the MaNGA, GALEX, and 
Very Large Array Sky Survey (VLASS) data
used for the study and the 
associated analysis. 
We discuss the possibility of AGN feedback
in NGC 3982 and provide a summary\ in Sect. 3.
We adopt a flat Universe cosmology with 
H$_{\rm{o}} = 71\, \mathrm{km\, 
s^{-1}\, Mpc^{-1}}$, $\Omega_{\rm{M}} = 0.27$, and 
$\Omega_{\Lambda} = 0.73$ \citep{komatsu2009five}.

\section{Data and analysis}

\begin{figure}
\centering
        \includegraphics[width=\columnwidth]{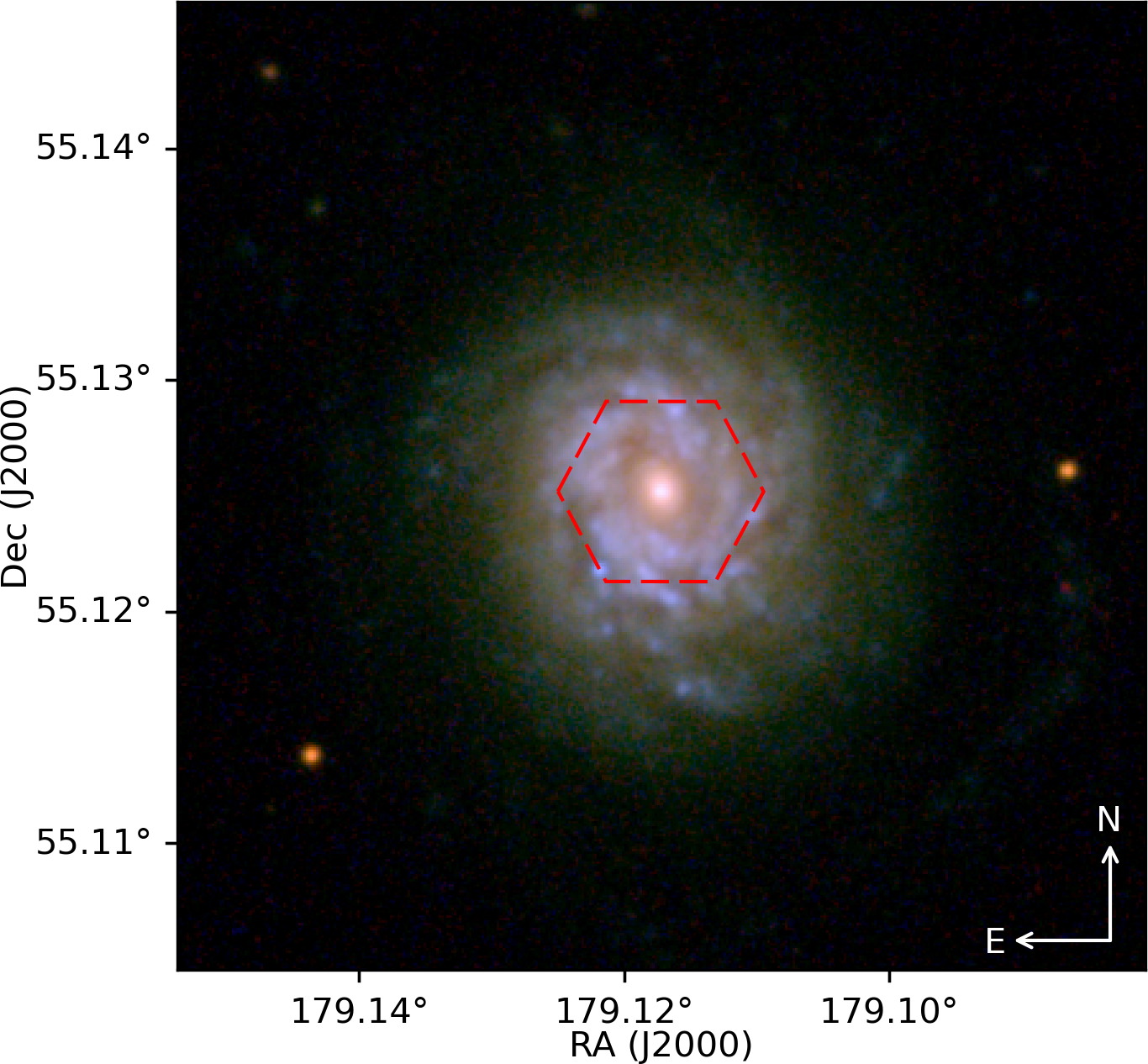}
    \caption{SDSS \textit{urz} colour image of NGC 3982. 
            The \textit{u} filter is shown in blue, \textit{r} 
            is in green, and \textit{z} is in red.
            We note the blue regions surrounding the relatively
            redder nuclear region.
             The MaNGA IFU hexagonal aperture 
             field of view is overlaid in red.
             }
    \label{fig:heaxagon}
\end{figure}

\begin{table}
\centering

\caption{General properties of NGC 3982.}
\label{tab:properties}
\resizebox{\columnwidth}{!}{%
\begin{tabular}{@{}lrl@{}}
\toprule
\multicolumn{1}{c}{Property} & \multicolumn{1}{c}{Value}     & \multicolumn{1}{c}{References}            \\ \midrule
Redshift (z)$^a$ & 0.00371  & \cite{accetta2022seventeenth} \\
Distance$^b$ & 15.6 Mpc  & \cite{wright2006cosmology} \\
Seyfert type                 & 1.9                           & \cite{veron2010catalogue}                     \\
L$_{\text{(2-10 KeV)}}$ &
  10$^{40-41}\ \text{erg}\ \text{s}^{-1}$ &
  \cite{kammoun2020hard} \\
L$_{\text{(5 GHz)}}$ &
  $\sim$10$^{36}\ \text{erg}\ \text{s}^{-1}$ &
  \cite{bontempi2012physical} \\
Black hole mass & $\sim$10$^{7}$ M$_{\sun}$ & \cite{beifiori2009upper}                    \\ \bottomrule
\end{tabular}%
}
\footnotesize{\raggedright (a) From SDSS optical spectra.\\ 
\raggedright (b) Derived from z. \\
}
\end{table}

\begin{figure}
\centering
        \includegraphics[width=\columnwidth]{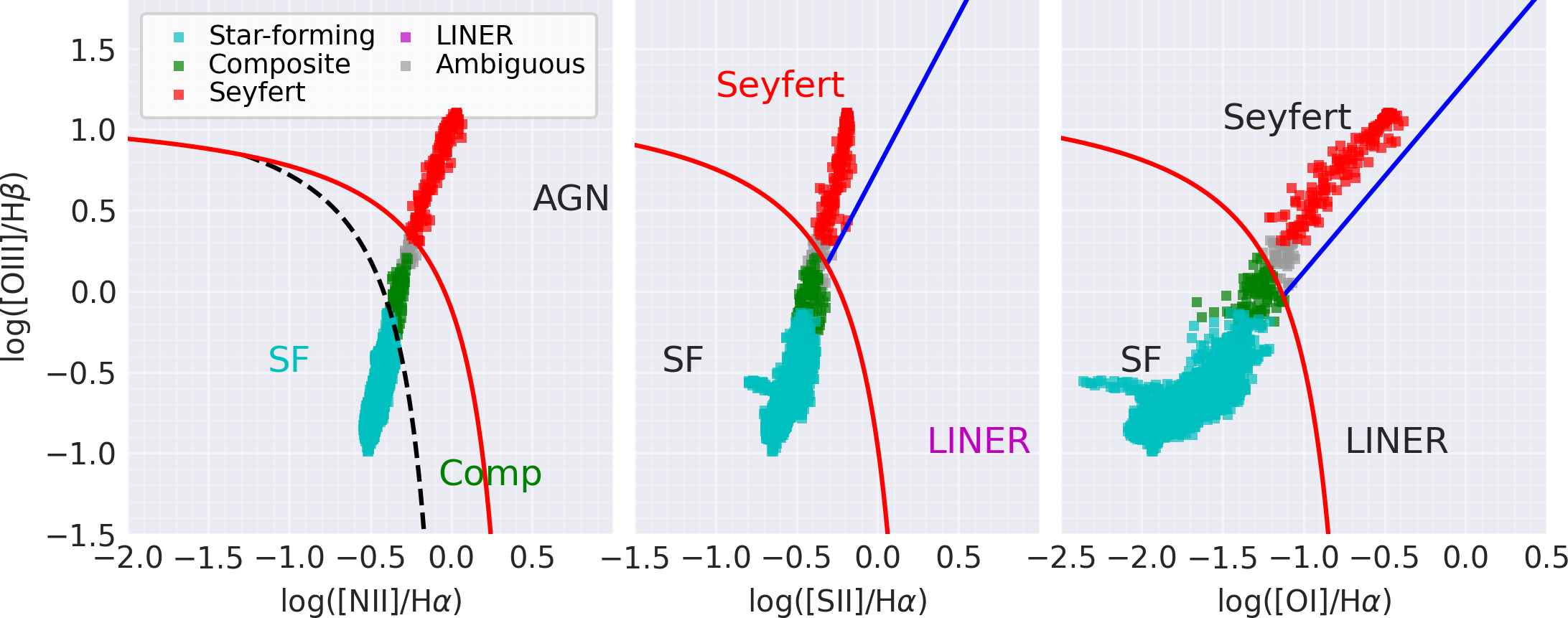}
    \caption{BPT diagram for NGC 3982 showing
             $[\text{O\ III}]/\text{H}_{\beta}$ versus
             $[\text{N\ II}]/\text{H}_{\alpha}$, $[\text{O\ III}]/\text{H}_{\beta}$ versus
             $[\text{S\ II}]/\text{H}_{\alpha}$, and $[\text{O\ III}]/\text{H}_{\beta}$ versus
             $[\text{O\ I}]/\text{H}_{\alpha}$
             plots.
             \textit{Left}: MaNGA spaxels classified into
             star-forming (SF), AGN, composite
             (AGN + SF), and ambiguous categories.
             \textit{Middle and Right}: MaNGA spaxels 
             categorised as SF, Seyfert, and LINER. 
             }
    \label{fig:bpt}
\centering
        \includegraphics[width=\columnwidth]{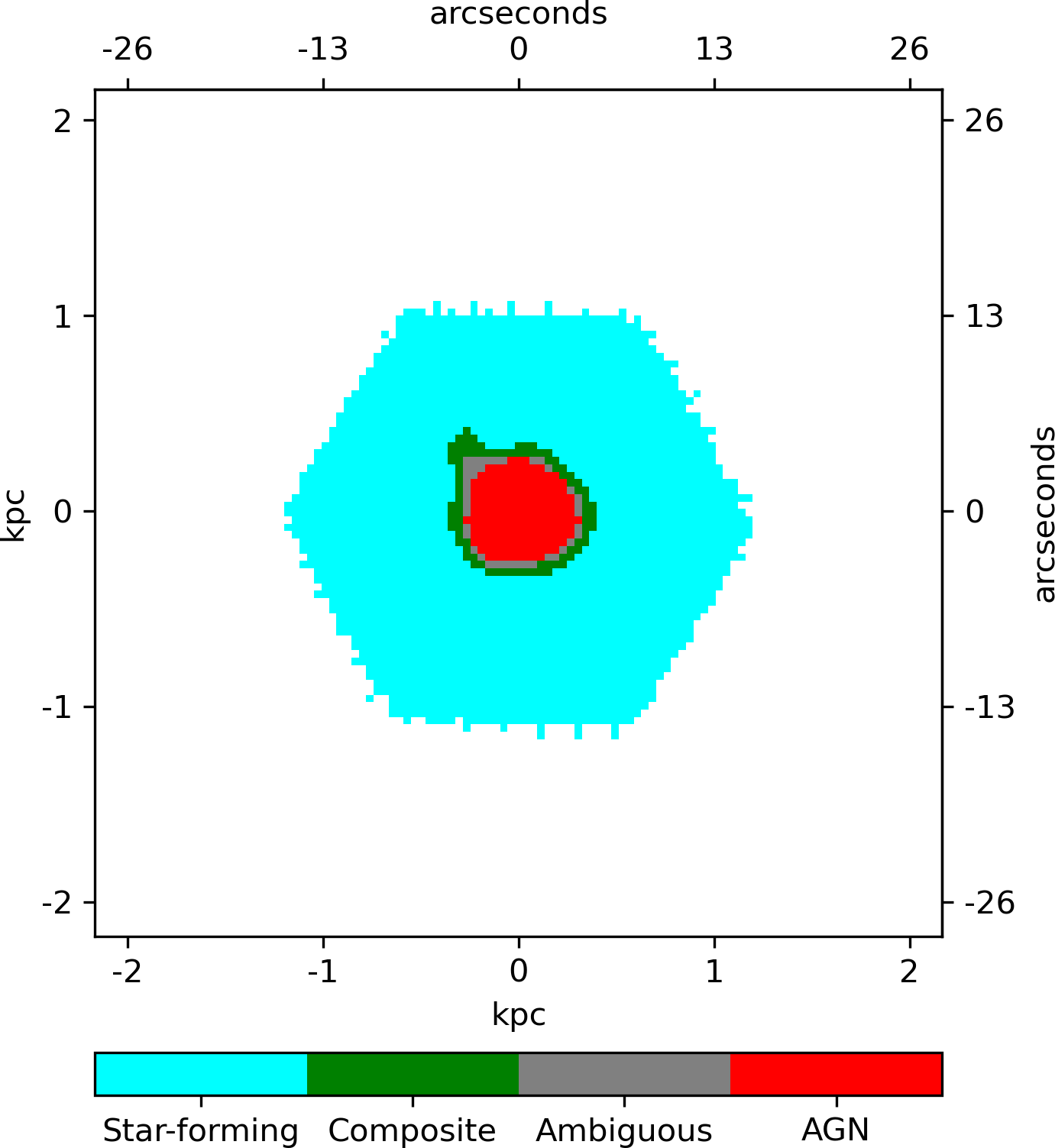}
    \caption{Spaxel BPT diagram classification of 
             Fig.~\ref{fig:bpt}.
             The figure origin coincides
             with the galaxy centre.
             North is up, and east is towards the
             left. 
             The angular offsets from the origin
             are given in the top and right
             axes.
             The angular offsets have been 
             converted to kiloparsec scales, shown             at the bottom and left axes.
             Star-forming (cyan), composite (green),
             ambiguous (grey), and AGN (red) categories
             are shown.}
    \label{fig:bpt_spaxel}
\end{figure}


The MaNGA survey observed galaxies 
up to redshift z $\sim$ 0.03.
There are 10010 unique galaxies  
in the final MaNGA data release, DR17 
\citep{accetta2022seventeenth}.
The reduced MaNGA data have a median 
angular resolution of 
2.54 arcseconds \citep{law2016drp}. 
A detailed description of the MaNGA sample design
can be found in \cite{wake2017sampledesign}.
NGC 3982 was observed 
with a 127-fibre IFU of 32 arcseconds diameter
as part of the MaNGA survey (Manga-ID: 1-189584). 
A summary of a few observed and derived properties
of the galaxy is given in Table~\ref{tab:properties}.
The SDSS \textit{urz} colour image cutout of NGC 3982 
with the MaNGA IFU hexagonal aperture is shown in
Fig.~\ref{fig:heaxagon}. 

GALEX was a UV survey mission that had near-ultraviolet 
(NUV; 5.3 arcseconds full width at half maximum)
and  far-ultraviolet (FUV; 4.2 arcseconds full width at half maximum) channels 
\citep{morrissey2007calibration}.
GALEX observed NGC 3982 in the NUV channel 
for an effective exposure time  of 
$\sim$2200 seconds
(GALEX tile: GI6\_012039\_HRS74\_75).
We accessed the NUV imaging data from the 
Mikulski Archive for Space Telescopes (MAST)
GALEX archive (no FUV data were available). 
The MaNGA resolution corresponds to 0.193 kpc, 
and the GALEX NUV resolution is 0.403 kpc at 
the galaxy's distance.


We checked the nature and extent of AGN 
ionisation in the central region of NGC 3982 as 
described below. 
The MaNGA data-analysis pipeline (DAP) 
generates secondary data products derived 
from MaNGA spectroscopy \citep{westfall2019dap}.
The stellar and emission-line 
kinematics are derived using penalised 
pixel-fitting (PPXF) software 
by simultaneously fitting a modified 
MILES stellar library and Gaussian emission-line 
templates to the MaNGA spectra (the PPXF 
software employs a maximum penalised likelihood approach
to fit templates to spectra; \citealt{cappellari2017ppxf, 
                  westfall2019dap, 
                  sanchez2006medium,
                  falcon2011updated}).
MaNGA DAP products are accessible through
\texttt{Marvin} software \citep{cherinka2019marvin}.
MaNGA DAP-generated maps of  
H$_{\alpha}$, 
H$_{\beta}$, 
$[\text{O\ III}] \ \lambda5007$, 
$[\text{O\ I}]\ \lambda6300$, 
$[\text{N\ II}]\ \lambda6584$, and 
$[\text{S\ II}]\ \lambda\lambda6717, 31$
emission line intensities were used in our subsequent analysis. 
To make a spaxel map of the present gas excitation 
mechanisms, 
we created a Baldwin-Phillips-Terlevich 
(BPT) diagram for NGC 3982  
using the \texttt{get\_bpt} function of \texttt{Marvin} 
\citep{baldwin1981classification}. 
\texttt{Marvin} uses the classification system 
of \cite{kewley2006host} 
and all three diagnostic criteria
from 
$[\text{O\ III}]/\text{H}_{\beta}$ versus
$[\text{N\ II}]/\text{H}_{\alpha}$, $[\text{O\ III}]/\text{H}_{\beta}$ versus
$[\text{S\ II}]/\text{H}_{\alpha}$, and $[\text{O\ III}]/\text{H}_{\beta}$ versus
$[\text{O\ I}]/\text{H}_{\alpha}$
plots.
It labels MaNGA spaxels as
star-forming (SF), AGN, or composite
(AGN + SF).
The function marks spaxels as ambiguous when it fails 
to classify them into one of these three categories. 
Seyfert galaxy (Seyfert) and low-ionisation narrow 
emission-line region (LINER) classification is also
carried out. 
A detailed explanation of the function
can be found on the \texttt{Marvin} 
documentation website\footnote{\url{https://sdss-marvin.readthedocs.io/en/latest/tools/bpt.html}}. 
The generated BPT diagram is shown in 
Fig.~\ref{fig:bpt}, and the on-sky map of 
spaxel classification in Fig.~\ref{fig:bpt_spaxel}.

\begin{figure}
\centering
        \includegraphics[width=\columnwidth]{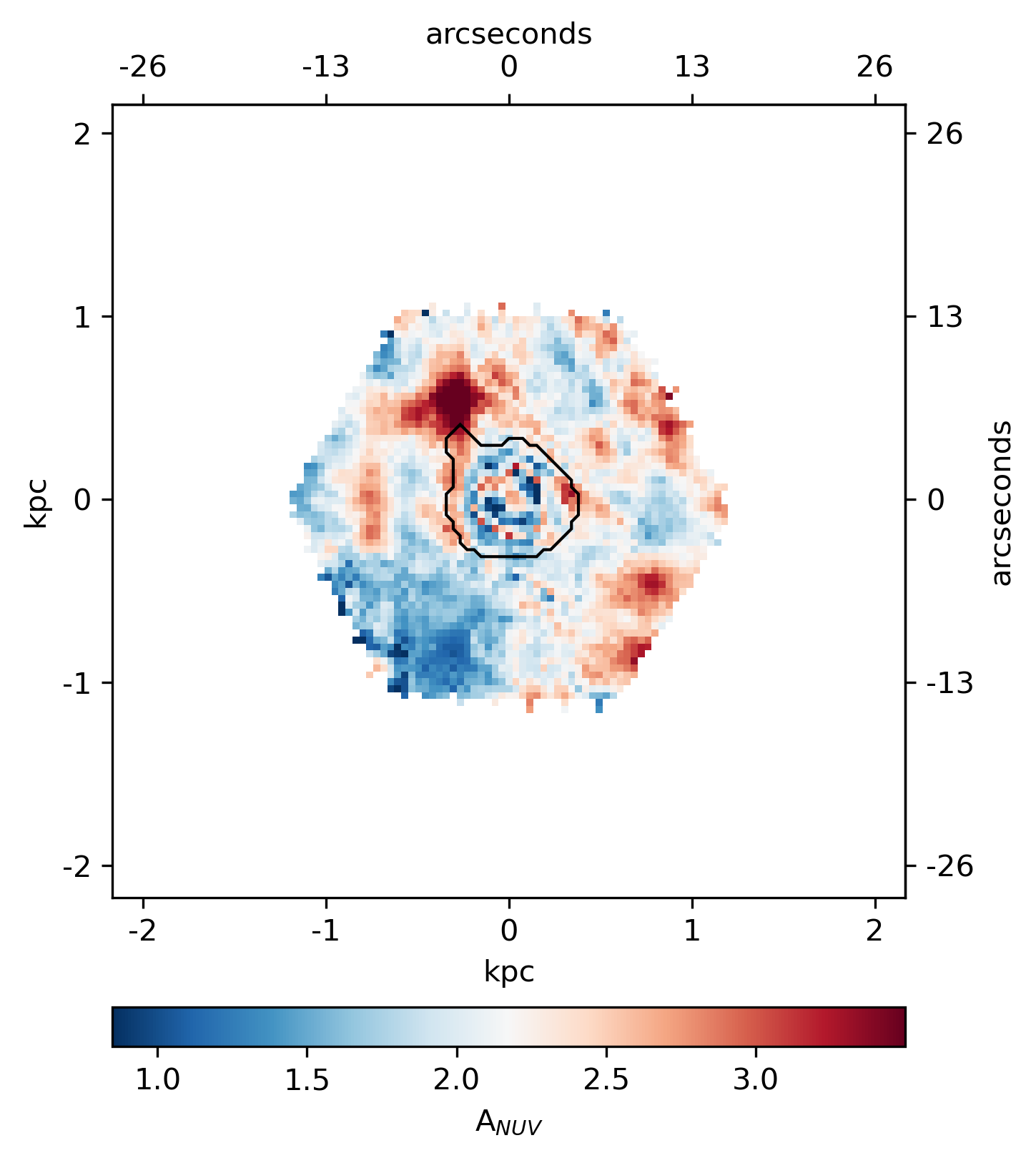}
    \caption{ MaNGA A$_{\text{NUV}}$ map 
             of NGC 3982.
             The black contour represents the area 
             encompassing the composite and AGN regions
in Fig.~\ref{fig:bpt_spaxel}.
             }
    \label{fig:A_NUV}
\end{figure}

We used the GALEX NUV band image to find the 
spatial distribution of recent star formation. 
The GALEX NUV image pixel units are in 
counts per second (CPS), which we converted 
to flux using the unit conversion factor given in \citet{morrissey2007calibration}: 
\begin{equation}
\text{F}_{\text{NUV}} =  2.06 \times 10^{-16} \times \text{CPS}
,\end{equation}
where $\text{F}_{\text{NUV}}$ is the NUV flux
in erg s$^{-1}$ cm$^{-2}$ \AA$^{-1}$.
The image in CPS units was subtracted 
for background before flux conversion.
The NUV flux was corrected for Galactic extinction 
by adopting a \cite{cardelli1989relationship} law 
with A$_{\text{V}}$ = 0.0437 \citep{schlegel1998maps} 
and R$_{\text{V}}$ = 3.1. 
The dust present in NGC 3982 can attenuate
the observed NUV band fluxes, leading
to inaccurate interpretation.
Therefore, we need to understand the spatial 
variation of dust attenuation levels, especially 
in the galaxy's central regions.
We used the Balmer decrement, the ratio between two 
H$\alpha$ and H$\beta$ emission line flux values, 
to estimate the dust attenuation.
We created the observed Balmer decrement map
by calculating the ratio between observed
$\text{H}_{\alpha}$ and $\text{H}_{\beta}$
emission line maps, 
$(\text{H}_{\alpha} / \text{H}_{\beta})_{\text{obs}}$.
Equation 4 from \cite{dominguez2013dust} 
was then used to convert the Balmer decrement 
to a colour excess map:
\begin{equation}
E(B-V) = 1.97\ \text{log}_{10} \left[ \frac{(\text{H}_{\alpha} / 
                                            \text{H}_{\beta})_{\text{obs}}}
                                          {(\text{H}_{\alpha} / 
                                            \text{H}_{\beta})_{\text{int}}} \right]
,\end{equation}
where $E(B-V)$ is the colour excess map and
$(\text{H}_{\alpha} / \text{H}_{\beta})_{\text{int}}$
is the expected Balmer decrement map without dust 
attenuation.
We used 
$(\text{H}_{\alpha} / 
  \text{H}_{\beta})_{\text{int}}$ = 3.1 
for spaxels falling inside
the AGN region of Fig.~\ref{fig:bpt_spaxel} 
and 2.86 for the remaining ones \citep{groves2012balmer}. 
Finally, we converted the colour excess map 
to an A$_{\text{NUV}}$ map (NUV band 
attenuation in magnitude) using a 
Calzetti attenuation law \citep{calzetti2000dust}:
\begin{equation}
\text{A}_{\text{NUV}} = k_{\text{NUV}} \times E(B-V)
,\end{equation}
where $k_{\text{NUV}}$ is the value on the
Calzetti reddening curve evaluated at 
the NUV effective wavelength (2315.7 $\AA$).
We find from the A$_{\text{NUV}}$ map 
(see Fig.~\ref{fig:A_NUV}) that  
the outer and central regions have comparable
attenuation levels, with a median A$_{\text{NUV}}$ 
value of 2.04 in the central AGN ionised regions 
and 2.15 in the galaxy disk.

To correct attenuation by dust in NGC 3982,
we used the A$_{\text{NUV}}$ map.
The extinction- and attenuation-corrected
NUV flux was calculated using
\begin{equation}
\text{F}_{\text{NUV, corrected}} = 
\text{F}_{\text{NUV}}\times
10^{0.4(\text{A}_{\text{NUV, Galactic}}\ 
+\ \text{A}_{\text{NUV}})}
,\end{equation}
where $\text{F}_{\text{NUV}}$ is the NUV flux,
$\text{A}_{\text{NUV, Galactic}}$ is the Galactic
extinction in the NUV band, and 
$\text{F}_{\text{NUV, corrected}}$ is the 
extinction- and attenuation-corrected NUV flux. 
The A$_{\text{NUV}}$ map created from the 
MaNGA data only covers the footprint shown in 
Fig.~\ref{fig:heaxagon}, 
which we need to extrapolate to the full galaxy 
to get the integrated extinction. 
Therefore, we used the median value of 
the A$_{\text{NUV}}$ map for the full 
extent of the NUV image of the galaxy. 
We note that this estimate should be 
considered a lower limit; it suggests that 
the flux could be attenuated by at least  a 
factor of 7.2.

\begin{figure}
\centering
        \includegraphics[width=\columnwidth]{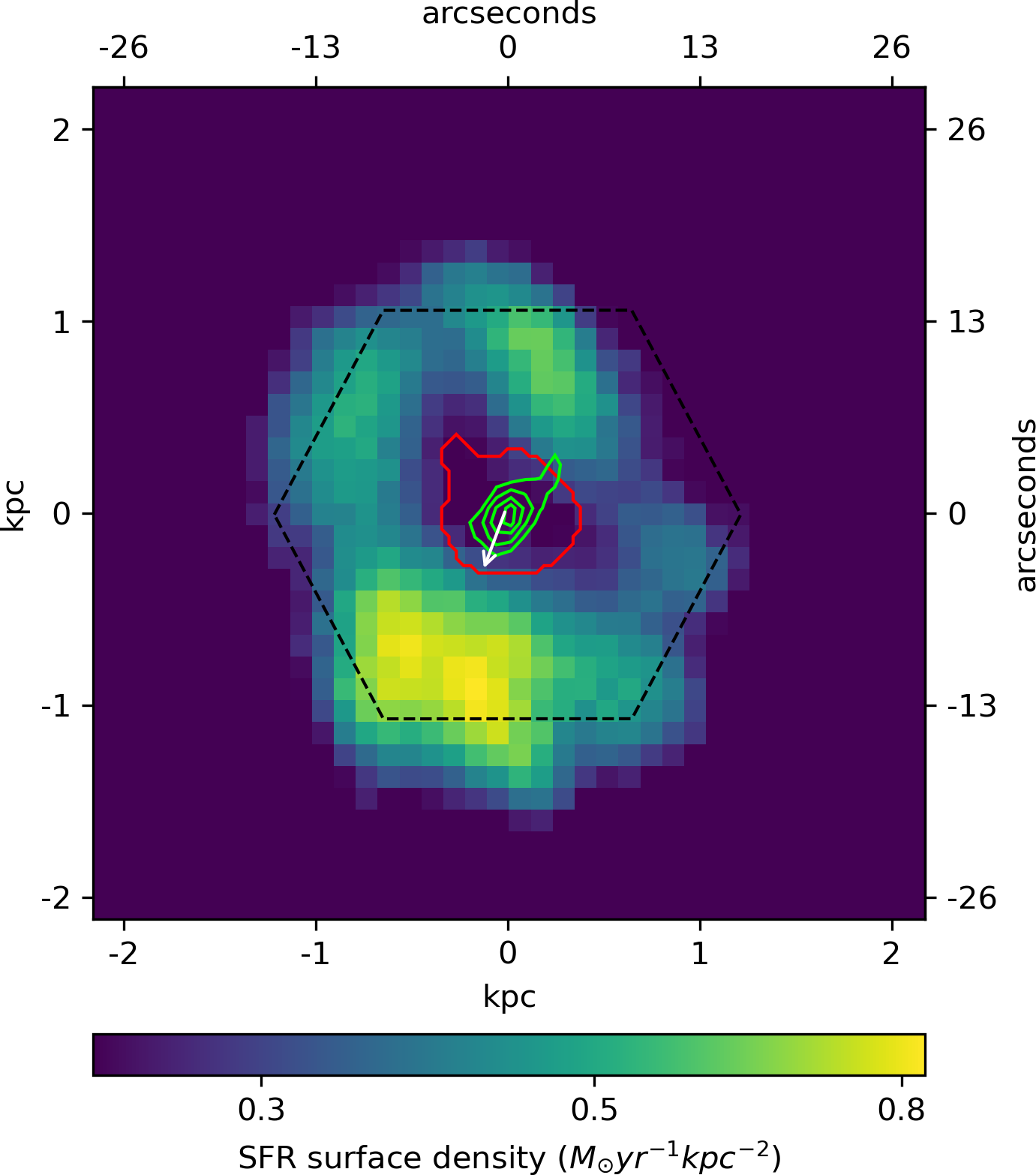}
    \caption{ GALEX NUV image of NGC 3982.
             The NUV image pixels in count rates
             have been converted to SFR surface densities 
             in $\text{M}_{\odot}\text{yr}^{-1}\text{kpc}^{-2}$.
             The red contour represents the area 
             encompassing the composite and AGN regions
             in Fig.~\ref{fig:bpt_spaxel}. The dashed black hexagon is the MaNGA IFU
             field of view.
             VLASS2.1 radio contours (green) with contour levels 
             0.0007, 0.0012, 0.0018, and 0.0024 Jy/beam 
             and a beamwidth of $\sim$2.5 arcseconds
             are overlaid on the figure. 
             The radio data reveal an elongated structure
             lying in the south-east to north-west direction.
             The VLBI-detected jet or outflow direction is 
             shown as a vector (white) for comparison.
             The vector has an origin at the flat spectrum core
             profile detected by \protect\cite{bontempi2012physical},
             and it points in the direction of
             their detected steep spectrum feature.
             }
    \label{fig:nuv}
\end{figure}

The NUV band luminosity, $\text{L}_\text{NUV}$, 
was calculated using\begin{equation}
\text{L}_\text{NUV} = 4\pi\ \text{D}^2\ \text{F}_{\text{NUV, corrected}}  
,\end{equation}
where D is the distance (see Table~\ref{tab:properties})
and $\text{F}_{\text{NUV, corrected}}$ is the attenuation-corrected flux. 
The NUV band luminosity is in erg s$^{-1}$.
The NUV luminosity was converted to the star
formation rate (SFR), assuming constant
star formation for 10$^8$ yr. 
Equation 4 from \cite{cortese2008ultraviolet}
was used to derive the SFR from $\text{L}_{\text{NUV}}$: 
\begin{equation}
\text{SFR}\ (\text{M}_{\odot}\text{yr}^{-1}) = 
\frac{\text{L}_\text{NUV}} {3.83 \times 10^{33}} \times 10^{-9.33}
.\end{equation}
Figure~\ref{fig:nuv} shows the NUV-derived 
SFR surface map of the galaxy.
A boundary contour encompassing the composite
and AGN regions from Fig.~\ref{fig:bpt_spaxel} 
is overlaid on the figure.
Also shown in the figure is the hexagonal 
aperture footprint of the MaNGA IFU. 
The cavity region in the centre matches 
the composite and AGN photoionised region.  
We estimated the median SFR density observed 
in the cavity and the ring-shaped
region; there is a factor of $\sim$2 reduction 
in the SFR density of the cavity region compared
to the ring-shaped region.

To check whether the lack of a full 
disk attenuation map of the galaxy affects our 
interpretation of the observed cavity region 
in the centre, we created an azimuthally 
averaged flux profile of the galaxy using 
the NUV image as follows. 
The axis ratio and position angle 
of NGC 3982 were found on the 
NASA/IPAC Extragalactic Database (NED) website\footnote{\url{https://ned.ipac.caltech.edu/byname?objname=ngc+3982}}.
Multiple elliptical apertures were defined
centred on NGC 3982 with an axis ratio 
of 0.901 and a position angle of 14.5$\degree$, 
each separated by 1.5 arcseconds along
the minor axis. 
The apertures were placed at up to $\sim$0.4 arcminutes (2 kpc),
and NUV fluxes were estimated. 
Then the differences between the aperture fluxes
were found to get the annuli fluxes.
Finally, annuli average fluxes were found by 
dividing by the annuli area. 
The annuli average flux is plotted as a function 
of distance from the galaxy centre in 
Fig.~\ref{fig:annuli_average}.
Similarly, elliptical apertures were used
on the A$_{\text{NUV}}$ map to estimate the
annuli average attenuation levels.
These levels were then used
to correct the annuli average flux values.
Attenuation-corrected annuli average fluxes are 
also plotted in Fig.~\ref{fig:annuli_average}.
We note that the A$_{\text{NUV}}$ map only covers
the central region of the NUV image.
Therefore, we can correct NUV fluxes
for attenuation  up to $\sim$0.2 arcminutes.
We stress that attenuation does not 
affect the NUV profile of the galaxy's 
central region, as demonstrated in  
Fig.~\ref{fig:annuli_average}.



We used Very Large Array Sky Survey (VLASS2.1) 
2-4 GHz radio sky survey data with an angular 
resolution of $\sim$2.5 arcseconds
\citep{lacy2020karl} to 
probe AGN activity in
NGC 3982.
We accessed the Quick Look images 
from the VLASS archive\footnote{\url{https://archive-new.nrao.edu/vlass/quicklook/}}. 
The VLASS contours are overlaid on the
NUV image in Fig.~\ref{fig:nuv}. 
The contours show an elongated structure 
lying in a south-east to north-west direction.
Interestingly, VLBI observations of NGC 3982 
in 1.7 and 5 GHz using the European VLBI Network reveal that there could be jet or outflow structures on 
milliarcsecond scales \citep{bontempi2012physical}.  
The VLBI-detected features are also oriented
in a south-east to north-west direction.

$[\text{O\ III}]$ is a forbidden optical 
emission line that is more excited by strong 
ionising sources, such as AGN, than star-forming
regions. 
Spatially resolved $[\text{O\ III}]$ flux 
and kinematics maps have been used in AGN 
feedback studies to trace ionised gas outflows 
(for example: \citealt{shin2019positive, ruschel2021agnifs}).
The $[\text{O\ III}]$ velocity dispersion in 
AGN host galaxies is mostly due to AGN activity 
\citep{rakshit2018census, woo2016prevalence}.
The MaNGA DAP maps of $[\text{O\ III}]$
flux and $[\text{O\ III}]$-traced ionised gas 
velocity dispersion are shown in Fig.~\ref{fig:oiii}.
The velocity dispersion map has been corrected for
instrumental dispersion.



\section{Discussion and summary}

\begin{figure}
\centering
        \includegraphics[width=\columnwidth]{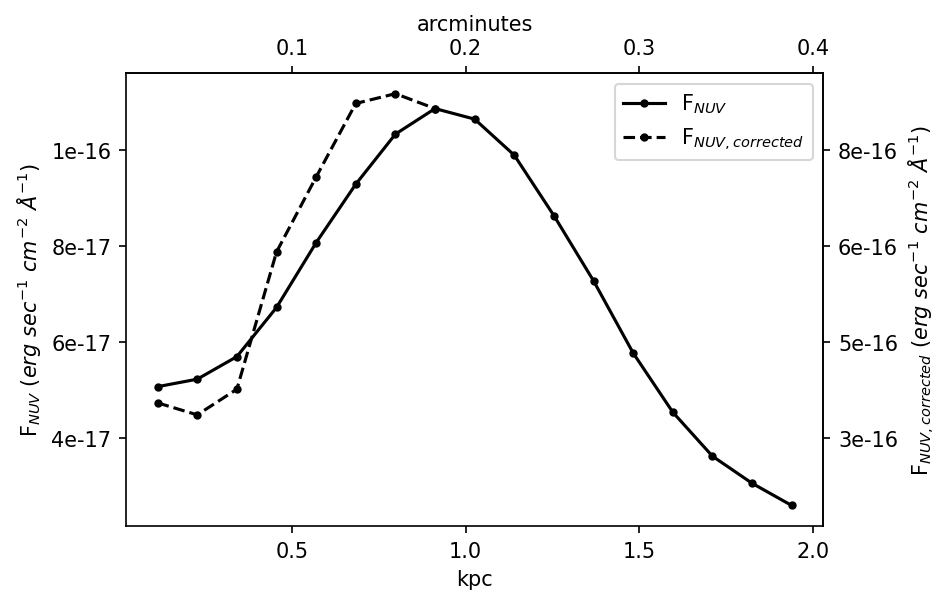}
    \caption{Azimuthally averaged NUV flux profile 
             of the galaxy. 
             Both attenuation-corrected and uncorrected 
             flux profiles are shown. 
             }
    \label{fig:annuli_average}
\end{figure}

\begin{figure*}
\centering
        \includegraphics[width=1.5\columnwidth]{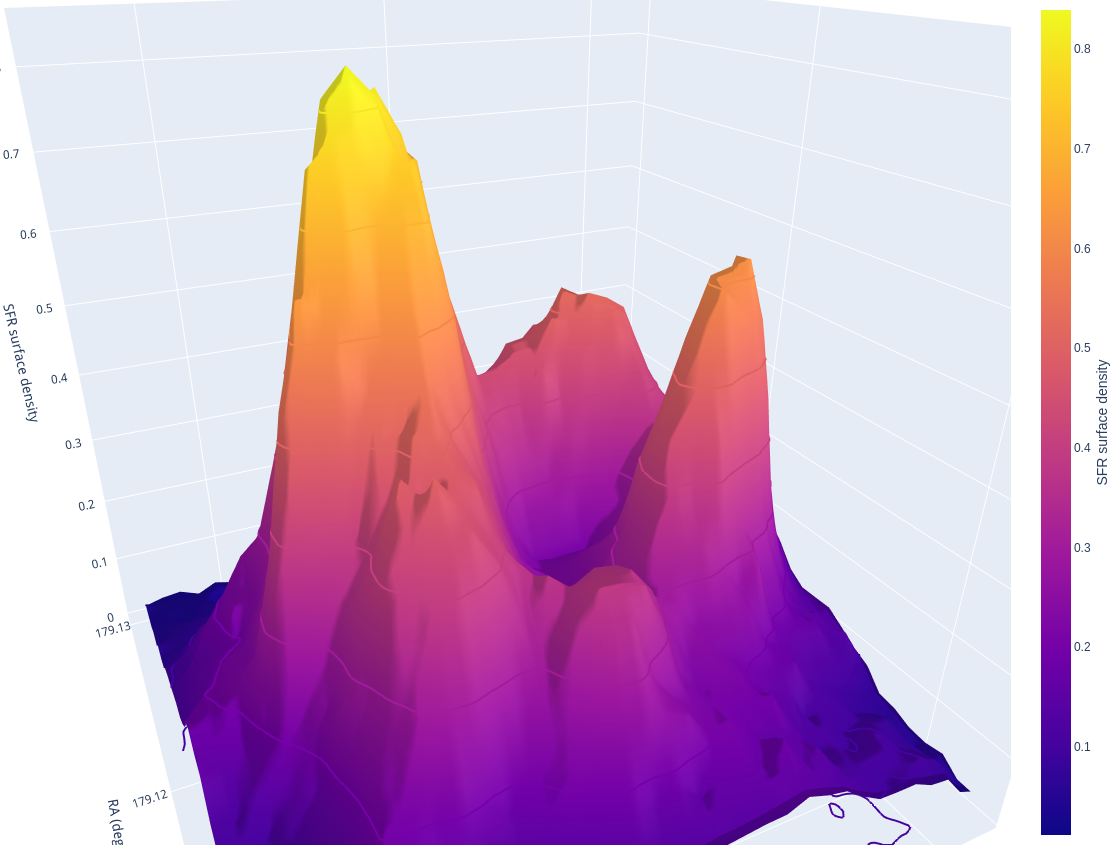}
    \caption{Snapshot from the interactive 
             3D visualisation of the SFR surface
             density profile of NGC 3982. 
             The visualisation is hosted at 
             \url{https://prajwel.github.io/NGC3982/}. 
             }
    \label{fig:3dplot}
\end{figure*}

 The NUV band directly traces stars formed
over the last 200 Myr and, therefore, probes 
recent star formation \citep{kennicutt2012star}.
The GALEX NUV-derived profile of SFR 
surface density shows recent star 
formation in a ring-like region around the 
centre of NGC 3982 (see Fig.~\ref{fig:nuv}).
We also created an interactive 
3D visualisation\footnote{The visualisation is hosted at \url{https://prajwel.github.io/NGC3982/}}
of the SFR surface density profile.
A snapshot from the interactive 
plot is shown in Fig.~\ref{fig:3dplot}.
It appears that star formation is suppressed in the 
central region of NGC 3982.

The intrinsic UV flux from star-forming 
regions can be modified due to dust attenuation. 
Even if the central region has the same 
levels of intrinsic NUV emission as the 
ring-like region, a large attenuation in 
the central region can produce the presently 
observed NUV profile.
However, the A$_{\text{NUV}}$ map shown
in Fig.~\ref{fig:A_NUV} has comparable 
dust attenuation levels in the central 
and outer regions, with a median A$_{\text{NUV}}$ 
value of 2.04 in the central AGN ionised regions 
and 2.15 in the galaxy disk. 
We do not see large attenuation levels 
in the central region.
Also, the $\text{F}_{\text{NUV, corrected}}$
profile of the galaxy closely matches the
$\text{F}_{\text{NUV}}$ profile 
(see Fig.~\ref{fig:annuli_average}).  
Therefore, dust attenuation of NUV flux 
in the galaxy cannot explain the cavity.


The observed suppression of star formation 
should be a real feature, not an artefact 
of attenuation.
This suggests that processes in the central 
region prevented star formation in the last 
200 Myr. 
The observed NUV cavity is approximately
shaped like an elongated ellipse. 
We estimate that it has major 
and minor axis lengths of $\sim$17 and $\sim$8 arcseconds,
respectively
($\sim$1.3/0.6 kpc).
The two distinct observational features 
that require explanation are the ring-like 
star-forming and star-formation-suppressed 
central regions.

The first possible explanation we considered 
is the presence of a bar. 
Bars are observed to induce star formation 
along the co-rotation radius and suppress star 
formation inside it \citep{george2020more}.
If present, a bar may produce the observed 
NUV profile in the galaxy.
However, \cite{regan1999using} studied the central 
region of NGC 3982 using the \textit{Hubble} Space Telescope and ruled out 
the presence of even a weak bar. 
NGC 3982 hosts an AGN, and the cavity region 
is covered by composite and AGN emissions. 
The boundary contour encompassing 
the composite and AGN regions
overlaid in Fig.~\ref{fig:nuv} shows the
extent of AGN ionisation in the galaxy
estimated using a BPT diagram.
Active galactic nuclei are known to suppress star formation 
and can also be associated with positive 
feedback.
Therefore, the likely mechanism for  
star formation suppression in NGC 3982 
is a jet or outflow associated with the AGN activity.

\begin{figure*}
\centering
        \includegraphics[width=2\columnwidth]{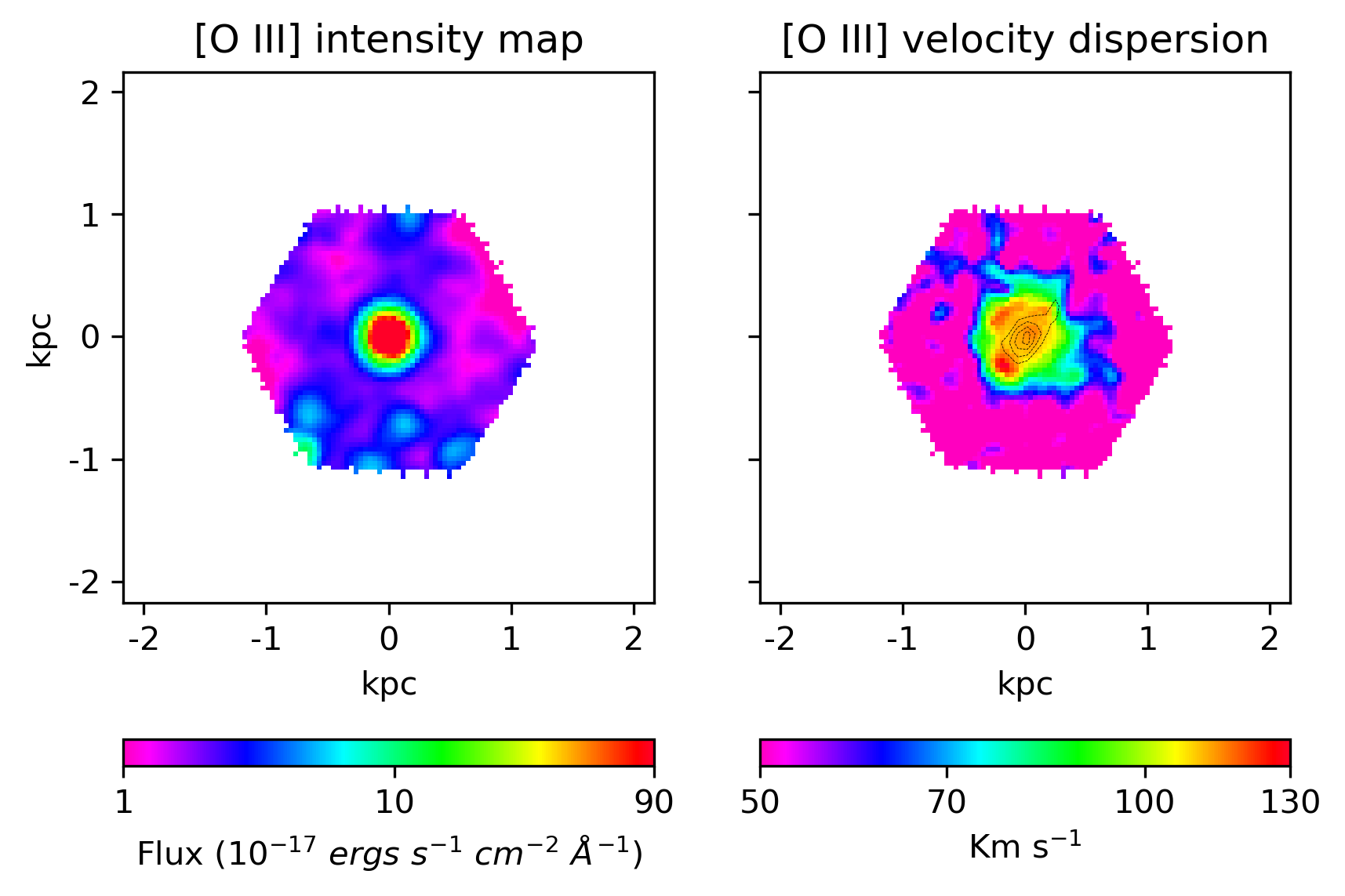}
    \caption{ MaNGA derived maps of $[\text{O\ III}]$
             emission line intensity and $[\text{O\ III}]$ traced 
             ionised gas velocity dispersion. 
             VLASS2.1 radio contours with contour levels 
             0.0007, 0.0012, 0.0018, and 0.0024 Jy/beam 
             and a beamwidth of $\sim$2.5 arcseconds
             are overlaid on the second panel in dashed
             black lines.}
    \label{fig:oiii}
\end{figure*}

The radio observations and $[\text{O\ III}]$ 
flux and velocity dispersion maps provide clues
regarding AGN activity.
The VLASS2.1 observations reveal an extended structure
with elongation in the south-east to north-west 
direction (see Fig.~\ref{fig:nuv}). 
While the $[\text{O\ III}]$ flux map shows 
a symmetrically distributed high 
emission at the central region, the velocity 
dispersion map shows signs of perturbations 
in the gas (see Fig.~\ref{fig:oiii}). 
The perturbations are most prominently observed
in the south-east ($\sim$130 km s$^{-1}$) and north-east 
($\sim$120 km s$^{-1}$) regions.
The perturbed gas  in the south-east is aligned with the
elongated VLASS radio structure, and the 
perturbed gas  in the north-east runs parallel to it.
These observations clearly show that 
the ionised gas is perturbed in the 
AGN-dominated region and that such perturbation 
may be driven by an AGN jet or outflow.
\cite{brum2017dusty}, who analysed the 
ionised gas kinematics, noted that a 
mild nuclear outflow could be present in the galaxy.
But compared to other local galaxies hosting
outflows, ionised gas outflow signatures in 
NGC 3982 are not prominent \citep{ruschel2021agnifs}.

The AGN hosted in NGC3982 is classified 
as Seyfert 1.9 and likely belongs 
to the radiative-mode AGN population.
Therefore, the galaxy may have an outflow 
driven by radiation.
We note that the AGN X-ray luminosity 
is $\sim$5 orders of magnitude greater than the 
radio luminosity (see Table~\ref{tab:properties}).
Nevertheless, the AGN luminosities in different bands 
may vary within a short time span (for example, 
3C 273; \citealt{soldi2008multiwavelength}). 
A jet may produce outflows with features 
similar to that generated by radiation 
\citep{cielo2018agn}. 
Regardless of the nature of the AGN activity,
NGC 3982 is perhaps one of the best 
examples of a nearby low-mass AGN  
that affects star formation in the disk.

It is interesting to compare
the NGC 3982 case with other observations 
of AGN feedback. 
Although the activity is presently not detectable, 
the AGN in NGC 7252 could have been active 
until recently \citep{schweizer2013iii} 
and suppressed star formation in the central 
region of the galaxy \citep{george2018uvit}.
In NGC 5728, a star-forming 
ring with a radius of $\sim$1 kpc  and a cavity is observed, and AGN 
outflow is found to be responsible 
\citep{shin2019positive}. 
Surprisingly, ring-like star formation is
also observed in NGC 7252 and NGC 3982 
at a radius of $\sim$1 kpc from the centre.
While all three galaxies may host a ring-like star-forming
region with a radius of  
$\sim$1 kpc attributed to AGN feedback, 
they differ in galaxy morphology and AGN
activity. 

Based on a multi-wavelength analysis, 
we present evidence for star 
formation suppression in the central 
region of the Seyfert galaxy NGC 3982.
As revealed from line diagnostic analysis, 
the galaxy's central region with reduced 
star formation is dominated by AGN+composite 
emission.
This is further supported by an AGN 
jet or outflow revealed from radio and 
gas velocity dispersion map analysis. 
The most plausible explanation for the observed 
scenario presented here is suppression of star 
formation in the central regions due to feedback 
from recent AGN activity.



\begin{acknowledgements}
 Astropy, IPython, Matplotlib, NumPy, Photutils, 
Plotly, Reproject,
SAOImage DS9, and SciServer were used for data analysis, 
viewing, and plotting 
\citep{astropy:2013, 
       astropy:2018, 
       ipython2007, 
       matplotlib, 
       numpy, 
       photutils, 
       plotly,
       robitaille2020reproject,
       ds9,
       taghizadeh2020sciserver}.
Funding for the Sloan Digital Sky 
Survey IV has been provided by the 
Alfred P. Sloan Foundation, the U.S. 
Department of Energy Office of 
Science, and the Participating 
Institutions. 

SDSS-IV acknowledges support and 
resources from the Center for High 
Performance Computing  at the 
University of Utah. The SDSS 
website is \url{www.sdss.org}.

SDSS-IV is managed by the 
Astrophysical Research Consortium 
for the Participating Institutions 
of the SDSS Collaboration including 
the Brazilian Participation Group, 
the Carnegie Institution for Science, 
Carnegie Mellon University, Center for 
Astrophysics | Harvard \& 
Smithsonian, the Chilean Participation 
Group, the French Participation Group, 
Instituto de Astrof\'isica de 
Canarias, The Johns Hopkins 
University, Kavli Institute for the 
Physics and Mathematics of the 
Universe (IPMU) / University of 
Tokyo, the Korean Participation Group, 
Lawrence Berkeley National Laboratory, 
Leibniz Institut f\"ur Astrophysik 
Potsdam (AIP),  Max-Planck-Institut 
f\"ur Astronomie (MPIA Heidelberg), 
Max-Planck-Institut f\"ur 
Astrophysik (MPA Garching), 
Max-Planck-Institut f\"ur 
Extraterrestrische Physik (MPE), 
National Astronomical Observatories of 
China, New Mexico State University, 
New York University, University of 
Notre Dame, Observat\'ario 
Nacional / MCTI, The Ohio State 
University, Pennsylvania State 
University, Shanghai 
Astronomical Observatory, United 
Kingdom Participation Group, 
Universidad Nacional Aut\'onoma 
de M\'exico, University of Arizona, 
University of Colorado Boulder, 
University of Oxford, University of 
Portsmouth, University of Utah, 
University of Virginia, University 
of Washington, University of 
Wisconsin, Vanderbilt University, 
and Yale University.
\end{acknowledgements}

\bibliographystyle{aa}
\bibliography{references}

\end{document}